\RequirePackage{lineno}
\documentclass[aps,prd,twocolumn,showpacs,amsmath,amssymb]{revtex4-1}
\usepackage{overpic}
\usepackage{graphicx}
\usepackage{epsfig}
\usepackage{dcolumn}
\usepackage{bm}
\usepackage{rotating}
\usepackage{multirow}
\usepackage{booktabs}
\usepackage{appendix}
\usepackage{subfigure}
\usepackage{hhline}
\usepackage{amssymb}
\usepackage{lineno}
\usepackage{hyperref}
\usepackage{mathrsfs}
\usepackage{verbatim}
\usepackage{amsmath}
\usepackage{amssymb}
\usepackage{amsfonts}
\usepackage{amsbsy}
\usepackage{float}
\usepackage{natbib}

\hypersetup{colorlinks,linkcolor=blue,anchorcolor=blue,citecolor=blue}
\include{def-com}

\def \half {{1\over 2}}

\def \cP {\mathcal{P}}
\def \cW {\mathcal{W}}

\newcommand{\eq}[1]{Eq.~\eqref{eq:#1}}
\newcommand{\eqs}[2]{Eqs.~\eqref{eq:#1} and \eqref{eq:#2}}

\hypersetup{colorlinks,linkcolor=blue,anchorcolor=blue,citecolor=blue}
\include{def-com}

\let\oldequation\equation
\let\oldendequation\endequation

\renewenvironment{equation}
  {\linenomathNonumbers\oldequation}
  {\oldendequation\endlinenomath}

\def \miss2{M_{\rm miss}^{2}}

\def \romanOne   {\uppercase\expandafter{\romannumeral1}}
\def \romanTwo   {\uppercase\expandafter{\romannumeral2}}
\def \romanThree {\uppercase\expandafter{\romannumeral3}}
\def \romanFour  {\uppercase\expandafter{\romannumeral4}}
\def \romanFive  {\uppercase\expandafter{\romannumeral5}}
\def \romanSix   {\uppercase\expandafter{\romannumeral6}}
\def \romanSeven {\uppercase\expandafter{\romannumeral7}}
\def \romanEight {\uppercase\expandafter{\romannumeral8}}
\def \romanNine {\uppercase\expandafter{\romannumeral9}}

\newcommand{\lambdacp}{\Lambda_{c}^{+}}

\newcommand{\sigmode}[1]{
	\ifnum#1=1
	\lambdacp \rightarrow n K_{S}^{0} \pi^{+}
	\else
	\ifnum#1=2
	\lambdacp \rightarrow n K_{S}^{0} K^{+}
	\fi
	\fi
}

\begin{document}

\DeclareGraphicsExtensions{.eps,.png,.ps}
 \newcommand{\BESIIIorcid}[1]{\href{https://orcid.org/#1}{\hspace*{0.1em}\raisebox{-0.45ex}{\includegraphics[width=1em]{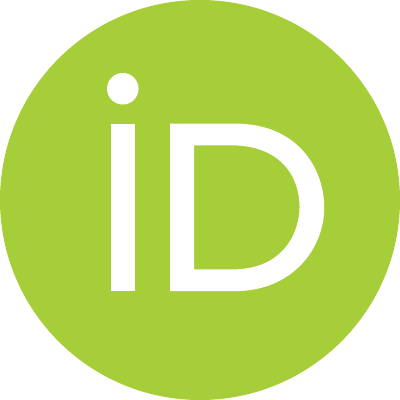}}}}

\title{\boldmath Phenomenological study of $\Omega_c\rightarrow  \Omega^-\pi^+$ at polarized electron-positron collider}

\author{Yunlu Wang}
\email{yunluwang20@fudan.edu.cn}
\affiliation{Key Laboratory of Nuclear Physics and Ion-beam Application (MOE) and Institute of Modern Physics, Fudan University, Shanghai, China 200433}

\author{Yunlong Xiao}
\email{xiaoyunlong@fudan.edu.cn(Corresponding author)}
\affiliation{Key Laboratory of Nuclear Physics and Ion-beam Application (MOE) and Institute of Modern Physics, Fudan University, Shanghai, China 200433}

\author{Pengcheng Hong}
\email{hongpc@ihep.ac.cn}
\affiliation{College of Physics, Jilin University, Changchun, China 130012}
\affiliation{Institute of High Energy Physics, Beijing 100049}

\date{\today}

\begin{abstract}

The exploration of symmetry laws stands as a cutting-edge direction in modern physics research. This work delves into the examination of $P$ and $CP$ symmetry properties within the charm quark system by analyzing asymmetry parameters in the two-body decay process of $\Omega_c$. By accounting for the polarization effects of electron and positron beams and employing the helicity formalism, we systematically analyze the decay characteristics of $\Omega_c$ and its subsequent hyperon decays through specific asymmetry parameters. A comprehensive formulation of the angular distribution for these decay processes has been developed. The research assesses the detection sensitivity of asymmetry parameters in the $\Omega_c\rightarrow  \Omega^-\pi^+$ decay mode across different experimental conditions, including varying data sample sizes and beam polarization configurations. These results contribute to enriching a theoretical foundation for forthcoming experimental endeavors at the STCF, offering significant implications for symmetry studies in the charm sector.

\end{abstract}

\maketitle
\oddsidemargin -0.2cm
\evensidemargin -0.2cm

\section{\boldmath Introduction}

The observed matter-antimatter asymmetry of the universe remains one of the most compelling puzzles in modern physics. While the Big Bang should have produced equal amounts of matter and antimatter, cosmological observations unequivocally show a universe dominated by matter. This indicates that there must be a mechanism in the evolution of the universe that favors matter over antimatter. Charge parity ($CP$) violation is one of Sakharov’s three essential conditions for understanding the matter and anti-matter asymmetry in the universe~\cite{Sakharov:1967dj}. 
In the Standard Model of particle physics, the quark dynamics are described by the Cabibbo-Kobayashi-Maskawa (CKM) mechanism~\cite{Cabibbo:1963yz,Kobayashi:1973fv}. 
This theoretical framework was experimentally confirmed through a series of landmark measurements. 
The BaBar and Belle collaborations reported the observation of indirect $CP$ violation in the decay of $B$ mesons~\cite{Belle:2001qdd,BaBar:2001ags} in 2001. Subsequently, the direct violation was observed during the rare decay process of $B$ mesons~\cite{Belle:2004nch,BaBar:2004gyj}. The first $CP$ violation in the charm system was observed by LHCb collaboration in 2019~\cite{LHCb:2019hro}. Recent LHCb results have significantly expanded the experimental landscape, with evidence of $CP$ violation in $\Lambda_b \to \Lambda K^+K^-$~\cite{LHCb:2024yzj} and the first observation in $\Lambda_b \to p K^-\pi^+\pi^-$ decays~\cite{LHCb:2025ray}. 
Similar searches of $\Lambda$, $\Sigma^+$, $\Xi^0$, and $\Xi^-$ decays have also been performed by BESIII collaboration~\cite{BESIII:2022qax,BESIII:2023sgt,BESIII:2025fre,BESIII:2023drj,BESIII:2023jhj,BESIII:2021ypr}.
However, all $CP$ measurements are not sufficient to explain the difference between matter and antimatter. 
The search for new $CP$-violating mechanisms has thus become a central focus in particle physics, driving both theoretical innovations and next-generation experiments~\cite{He:2022bbs}.

Theoretically, the decay amplitude of $\Omega_c$ consists of factorizable and non-factorizable contributions. While non-factorizable contributions are typically negligible in charmed meson decays, $\Omega_c$ decays exhibit unique dynamics, as their amplitudes can receive contributions from both $W$-emission and $W$-exchange diagrams. The $W$-exchange diagram (manifesting as a pole diagram) escapes helicity and color suppression, leading to non-trivial contributions.
In the case of $W$-exchange processes, which can be represented as pole diagrams, the typical suppression mechanisms related to helicity and color no longer apply~\cite{Cheng:1993gf}. As a result, theoretical predictions for charmed baryon decays tend to be more complex compared to those for charmed mesons. Various theoretical frameworks have been developed to study these decays, such as the covariant confined quark model~\cite{Korner:1978ec,Korner:1992wi, Ivanov:1997ra}, the pole model~\cite{Cheng:1993gf, Cheng:1991sn,Sharma:1998rd}, and current algebra~\cite{Cheng:1993gf, Sharma:1998rd, Uppal:1994pt}.
The Cabibbo-favored decay $\Omega_c \rightarrow \Omega^- \pi^+$ is dominated by the $W$-emission process and also has a relatively large branching fraction, which makes it particularly interesting for studying parity ($P$) and $CP$ violation.
In $\Omega_c$ and $\bar\Omega_c$ two-body decays, the asymmetry parameters $\alpha_{\Omega_c}$ and $\alpha_{\bar{\Omega}_c}$ quantify the contributions of $P$-wave (parity-violating) and $D$-wave (parity-conserving) amplitudes~\cite{Lee:1957qs}. Specifically, $\alpha_{\Omega_c}$ and $\alpha_{\bar{\Omega}_c}$ parameterize the decays of baryons and anti-baryons, where non-zero $\alpha_{\Omega_c}$( $\alpha_{\bar{\Omega}_c}$) and $(\alpha_{\Omega_c} + \alpha_{\bar{\Omega}_c}) / (\alpha_{\Omega_c} - \alpha_{\bar{\Omega}_c})$ values mean $P$ and $CP$ violations.
These features make $\Omega_c$ decays an ideal laboratory for symmetry test and polarization study, which can be explored at next-generation facilities. The Super Tau-Charm Facility (STCF) is designed to achieve a peak luminosity of $10^{35}$ cm$^{-2}$s$^{-1}$~\cite{Achasov:2023gey}, representing a two-order-of-magnitude improvement over BESIII. With a center-of-mass energy range spanning $2-7$ GeV, which covers the production threshold for $\Omega_c\bar{\Omega}_c$ baryon pairs, STCF will enable unprecedented statistics for studies of charmed baryons. Recent theoretical studies have demonstrated that transverse and longitudinal beam polarization can improve parameter measurement precision at electron-positron colliders~\cite{Zeng:2023wqw,Cao:2024tvz}. 

In this work, we present a systematic investigation of how beam polarization affects $\Omega_c$ polarization by comparing with polar and azimuthal angle distributions of particles from $\Omega_c$ decay. We provide the first study of parameter sensitivity under different transverse and longitudinal polarization configurations. Furthermore, we evaluate the optimization effects of beam polarization schemes on $CP$ violation. These phenomenological results establish crucial theoretical support to understand $\Omega_c$ decay at future STCF.

\section{ANGULAR PARAMETERS}
\begin{figure}[thb]
\begin{center}
\includegraphics[scale=.4]{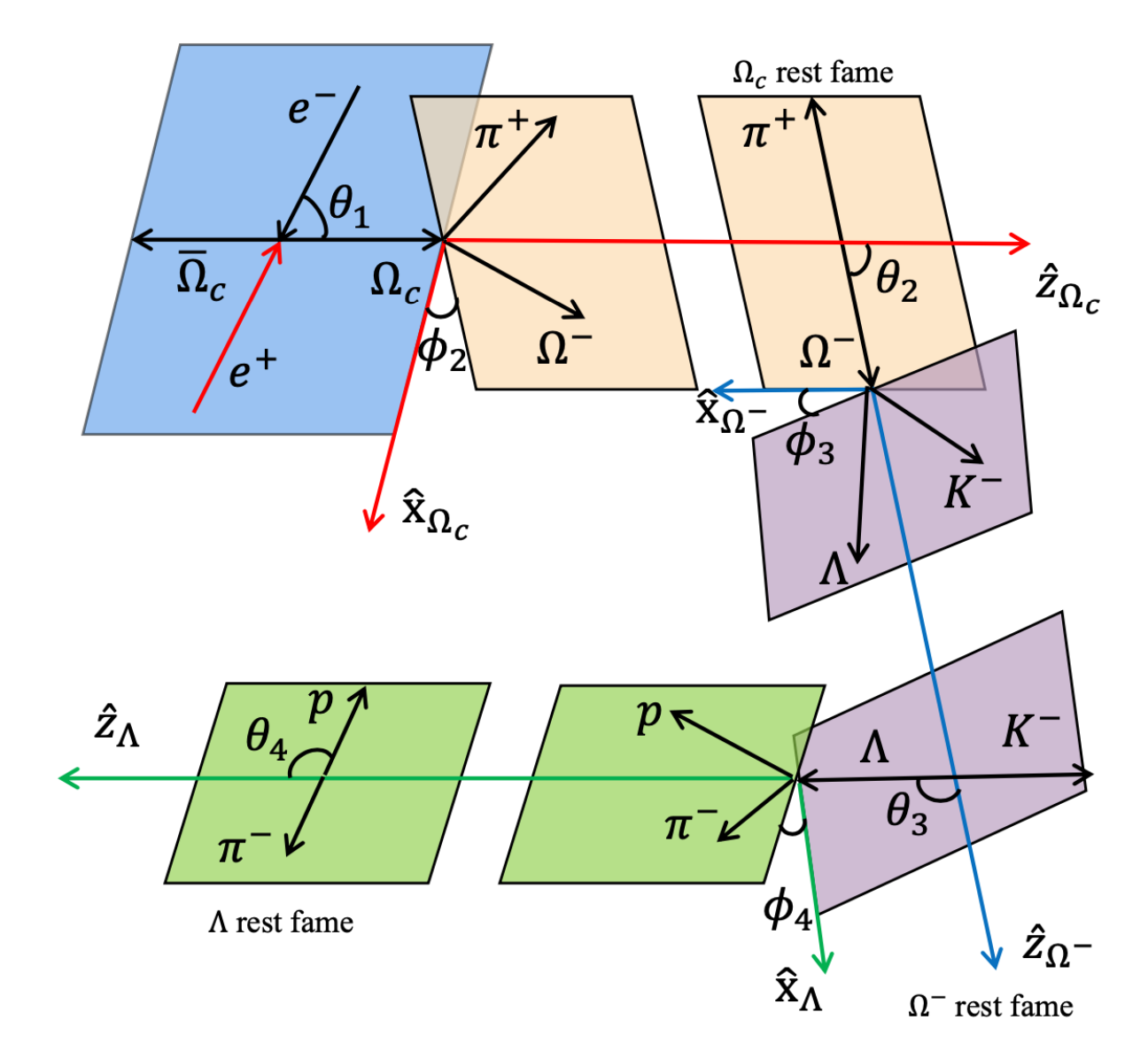}
\end{center}\vspace{-5ex}
\caption{Definition of helicity angles at $e^+e^-$ collider.}
\label{fig:angle}
\end{figure}

The decays in this analysis are described in a helicity formalism~\cite{{0025}}.
The helicity angles and amplitudes of the $\Omega_{c}\bar{\Omega}_{c}$ production and the decay of $\Omega_c$ are listed in Tab.~\ref{tab:decay}, while the corresponding angles are shown in Fig.~\ref{fig:angle}. The momenta $p_i$ are obtained by boosting particle $i$ to the rest frame of its mother particle, and the values of angles can be constructed by the momenta. 
 \begin{table}[htbp]
\caption{Helicity angles and amplitudes in relative decays.} 
\label{tab:decay}
\begin{tabular}{ccc}
\hline\hline
decay & helicity angle &helicity amplitude
\\\hline
\vspace{-3mm}
\\
$\gamma^*\rightarrow \Omega_c(\lambda_1)  \bar{\Omega}_c(\lambda_0)$ & $(\theta_1, \phi_1)$ & $A_{\lambda_1,\lambda_0}$   \\
$\Omega_c(\lambda_1)\rightarrow \Omega^-(\lambda_2)  \pi^+$ & $(\theta_2,\phi_2)$ & $B_{ \lambda_2 }$   \\
$\Omega^-(\lambda_2)\rightarrow \Lambda(\lambda_3)  K^-$ & $(\theta_3, \phi_3)$ & $F_{ \lambda_3 }$   \\
$\Lambda(\lambda_3)\rightarrow p(\lambda_4)  \pi^-$ & $(\theta_4, \phi_4)$ & $H_{ \lambda_4 }$   \\
 \hline\hline
\end{tabular}
\end{table}

In general, a helicity amplitude for a two-body decay process can be denoted by $A^J_{\lambda_i,\lambda_j}$, where $J$ denotes the spin of the mother particle, and $\lambda_i,  \lambda_j$ are daughter particles' helicity values which are the projection of particle spin in momentum direction. Under parity conservation, the amplitudes will obey
\begin{align}
A^J_{\lambda_i,\lambda_j}
=
\eta \eta_i \eta_j(-1)^{J-s_i-s_j} A^J_{-\lambda_i,-\lambda_j}
\,,
\end{align}
where $\eta_i$ is the intrinsic parity of each particle. The spins of the two daughter particles are denoted by $s_i, s_j$ respectively. 

For the first process in Tab.~\ref{tab:decay}, the helicity values can be $\lambda_1=\pm\frac12$ and $\lambda_0=\pm\frac12$ for spin-$\frac12$ particles $\Omega_c$ and $\bar{\Omega}_c$. The photon has spin $J=1$ and parity $\eta=1$, while the charmed baryons have $\eta_1=\eta_0=1$ and $s_1=s_0=\frac12$. For simplification, the superscript $J$ is suppressed in the following so that $A_{\lambda_i,\lambda_j}=A^J_{\lambda_i,\lambda_j}$. The other amplitudes $B, F$ and $H$ listed in Tab.~\ref{tab:decay} follow a similar notation. We also suppress the subscripts corresponding to the daughter particles $\pi^\pm$ and $K^-$ such as $B_{\lambda_2}=B_{\lambda_2,0}$, since they are spin-zero particles.

If all processes in Tab.~\ref{tab:decay} are parity conserved, the corresponding helicity amplitudes, $A,\, B,\, F,\, H$ satisfy following relations:
\begin{align}
A_{\frac12,\frac12} 
  = & A_{-\frac12,-\frac12}
\,, 
A_{\frac12,-\frac12} 
  =  A_{-\frac12,\frac12}
\,,\nonumber \\
B_{\frac12} 
  = & - B_{-\frac12}
\,, 
F_{\frac12} 
  =  - F_{-\frac12}
\,,
H_{\frac12} 
  =  - H_{-\frac12}\,.
\end{align}
Here, the scripts are suppressed if they are fixed constants. Specifically, since angular momentum conservation, only helicity values $\lambda_2=\pm\frac12$ of $\Omega^-$ are allowed, which leads to two amplitudes $B_{\pm\frac12}$. In Standard Model, the $P$ violations in weak decays make the above equations invalid. To quantify the violations, we define the following three asymmetric parameters:
\begin{align}
\alpha_{\Omega_c} 
\equiv&
{|B_{\half}|^2-|B_{-\half}|^2\over |B_{\half}|^2 +|B_{-\half}|^2}
\,,\\
\alpha_{\Omega^-} 
\equiv&
{|F_{\half}|^2-|F_{-\half}|^2\over |F_{\half}|^2 +|F_{-\half}|^2}
\,,\\
\alpha_{\Lambda} 
\equiv&
{|H_{\half}|^2-|H_{-\half}|^2\over |H_{\half}|^2 +|H_{-\half}|^2}\,.
\end{align}
They are equivalent to original definitions from Lee-Yang~\cite{Lee:1957qs} by decomposing amplitudes into $S$ and $P$ waves. The measured values of the last two parameters are $\alpha_{\Omega^-}=0.0154\pm 0.002$~\cite{HyperCP:2005wwr} and $\alpha_{\Lambda}=0.746\pm0.008$~\cite{BESIII:2024nif}, while the value of the parity parameter $\Omega_c\rightarrow \Omega^- \pi^+$ is still lack.

\begin{figure}[thb]
	\begin{center}
\includegraphics[scale=.55]{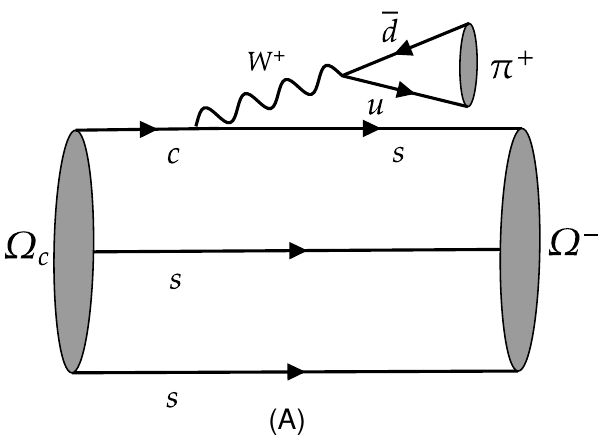}
\includegraphics[scale=.55]{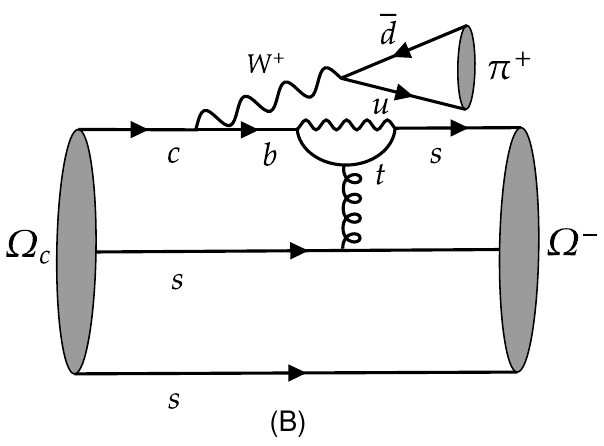}
	\end{center}\vspace{-5ex}
	\caption{Tree and penguin diagrams as examples for $\Omega_c\rightarrow\Omega^-\pi^+$ decay. }
	\label{fig:tree-loop}
\end{figure}

Moreover, if $CP$ conservation holds in charge conjugate decay $\bar{\Omega}_c\rightarrow \Omega^+\pi^-$, the parity parameter of this decay is expected to have the same absolute value but an opposite sign, like  $\alpha_{\bar{\Omega}_c} = - \alpha_{\Omega_c} $. Thus the $CP$ violation can be characterized by 
\begin{align}\label{eq:Acp}
\mathcal{A}_{CP}
\equiv &
\frac{ \alpha_{\Omega_c} + \alpha_{\bar{\Omega}_c}  }
{ \alpha_{\Omega_c} - \alpha_{\bar{\Omega}_c}  } \,.
\end{align}
This definition has an advantage that the systematic uncertainties of production and detection asymmetries are largely cancelled~\cite{Wang:2022tcm}. The approximation of the parameter can be written by ~\cite{BESIII:2021ypr,Donoghue:1986hh}
\begin{align}
\mathcal{A}_{CP}
\propto
-\tan\Delta\delta \tan\Delta\phi ,
\end{align}
after inserting partial amplitudes. The relative strong phase $\Delta\delta$ predominantly arises from final-state interactions, which are discussed and computed within QCD factorization framework~\cite{Beneke:2000ry,Beneke:2001ev} and resonance-induced model~\cite{Cheng:2004ru} for $B$ decays. The weak phase $\Delta\phi$ stems from interference between amplitudes with different partial wave configurations in the same decay~\cite{Saur:2020rgd}. The amplitudes partially derive from tree and penguin diagrams like the examples shown in Fig.~\ref{fig:tree-loop}, where the tree diagram is Cabibbo-favored and the penguin diagram is suppressed from off-diagonal elements $V^*_{cb}V^*_{ts}$. The magnitude of the weak phase shift is determined by $\text{Im}[-(V^*_{cb}V_{tb}V^*_{ts}V_{ud})]$ and expected at order of  $\mathcal{O}(10^{-4}\!\sim\!10^{-5}) $, which is close to predicted $CP$ violation of hyperons systems~\cite{Donoghue:1986hh,Tandean:2002vy}. However, an additional suppression of order $10^{-5}$ arises from the extra Fermi constant $G_F$ in the penguin diagram when the partial waves in Ref.~\cite{Wang:2024qff} are incorporated. Consequently, the observable $CP$ violation in \eq{Acp} is subject to further suppression. Including the direct $CP$ violation, a greater number cannot be ruled out for the mixing of strong violation and even new physics, that demands performing measurements at future $e^+e^-$ collider.

\section{Spin density matrix and angular distribution}

\subsection{$\Omega_c$ production at $e^+e^-$ collider}

The density matrix (SDM)~\cite{Doncel:1973sg,Chen:2020pia} involves the spin and polarization information of particles. The SDM of spin-$\frac12$ particles like $\Omega_c$ is expressed by a $2\times2$ matrix:
\begin{align}\label{eq:rho-Omec}
\rho^{\Omega_c}
= &
\frac{\cP_0}{2}
\begin{pmatrix}
1 + \cP_z & \cP_x - i\cP_y \\
\cP_x + i\cP_y & 1 - \cP_z
\end{pmatrix}\,,
\end{align}
where $\cP_0$ carries the unpolarized information, while the information of transverse polarization is taken by $\cP_x$ and $\cP_y$. Operator $\cP_z$ is longitudinal polarization. The information of initial collision is included by SDM of the virtual photon $\rho^{\gamma^*}$, which connects to $\rho^{\Omega_c}$ via the equation:
\begin{eqnarray}\label{eq:Xi-rho}
\rho^{\Omega_c}_{\lambda_1,\lambda_1^{'}}&=&\sum_{{\lambda},\lambda_0}{\rho^{\gamma^*}_{\lambda,\lambda^{'}}}D^{1*}_{\lambda,\lambda_{1}-\lambda_0}(\phi_1,\theta_1,0)\nonumber\\
	&\times&D^{{1}}_{\lambda',\lambda_{1}^{'}-\lambda_0}(\phi_1,\theta_1,0)A_{\lambda_{1},\lambda_0}A_{\lambda_{1}^{'},\lambda_0}^{*},
\end{eqnarray}
where $D^J_{\lambda_j,\lambda_k}$ is the Wigner-D function. For polarized symmetric $e^+e^-$ beams, when the transverse and longitudinal  polarizations are considered simultaneously, the SDM of polarized photon is given by~\cite{Cao:2024tvz}
\begin{eqnarray}
\rho^{\gamma^*}_{\lambda,\lambda'}
\!= &\!
\frac12\! \begin{pmatrix}\!
 (1-p_L)(1+\bar{p}_L) & 0 & p_T^2\\
 0 & 0 & 0 \\
 p_T^2 & 0 & (1+p_L)(1-\bar{p}_L)\!
\end{pmatrix}\!.
\end{eqnarray}
In a positron-electron annihilation experiment with symmetric beam energy, the electron and positron share the same transverse polarization $p_T$ respecting the $z$ axis, which is chosen along the positron momentum shown in Fig.~\ref{fig:angle}. The longitudinal polarizations of electron is represented by factor $p_L$, and the ratio of the number of right-handed and left-handed electrons in the beam is determined by the factor through $(1+p_L)/(1-p_L)$, while another factor $\bar{p}_L$ is for positron. The diagonal elements indicate the fact that the photon couples either a single right-handed particle to a left-handed antiparticle or a single left-handed particle to a right-
handed antiparticle.

We clarify the magnitude and phase angle of an amplitude by denoting $A_{\lambda_1,\lambda_0}=a_{1}e^{i\zeta_1}$ and define the phase difference $\Delta_1=\zeta_1-\zeta'_1$. The other amplitudes take similar decompositions. Substituting \eq{Xi-rho} into \eq{rho-Omec} under parity conservation of $\Omega_c$ production, we get the unpolarized section which depends on the polarizations of the virtual photon, that is 
\begin{align}
\label{eq:P0-photon}
\cP_0 = (1-p_L\bar{p}_L)(1 + \alpha_c\cos^2\theta_1) + p_T^2 \,\alpha_c  \sin^2\theta_1
\cos2\phi_1\,.
\end{align}
Here constant $\tfrac12|A_{\frac12,-\frac12}|^2+|A_{\frac12,\frac12}|^2$ is suppressed, since it does not contribute to the final cross sections after normalization. The angular distribution parameter $\alpha_c$ for this production is defined by
\begin{align}
	\alpha_c\equiv{|A_{\half,-\half}|^2-2|A_{\half,\half}|^2\over |A_{\half,-\half}|^2+2|A_{\half,\half}|^2}\,,
\end{align}
that has not been measured yet. In the following content, since the electromagnetic interaction
is dominated for the production of the charged pair, we use hypothetical values $\alpha_c = 0.7$ and $\Delta_1=\pi/6$, which are close to the measurements of strange baryons~\cite{BESIII:2016nix}. Our analysis framework is applicable to any value of $\alpha_c$, and the actual value can be measured via the angular distribution of event number given in \eq{N-theta1}.

\begin{figure}[thb]
	\begin{center}
	\includegraphics[scale=.60]{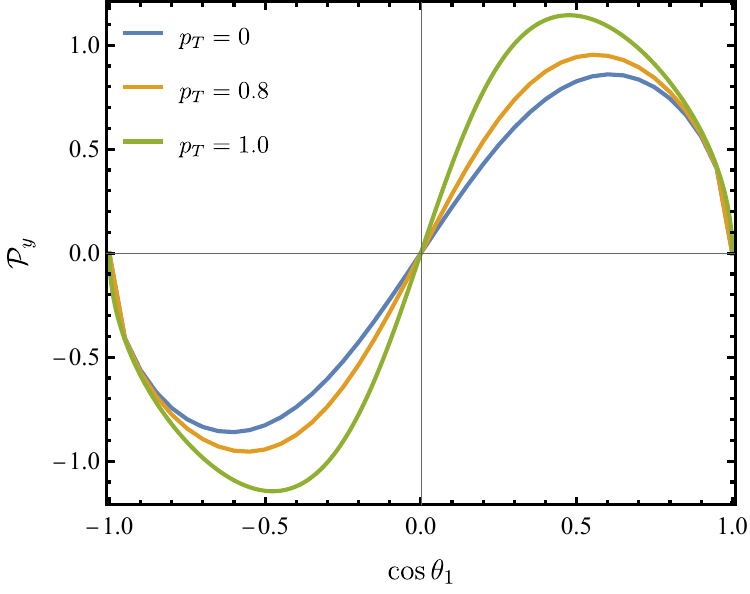}
	\end{center}\vspace{-5ex}
	\caption{The $\cos\theta_1$ distributions of polarization $\cP_y$ for different transverse polarizations of beams. Here $p_L=0$ is used.  }
	\label{fig:Py-theta1}
    \vspace{-5mm}
\end{figure}

\begin{figure}[thb]
	\begin{center}
	\includegraphics[scale=.60]{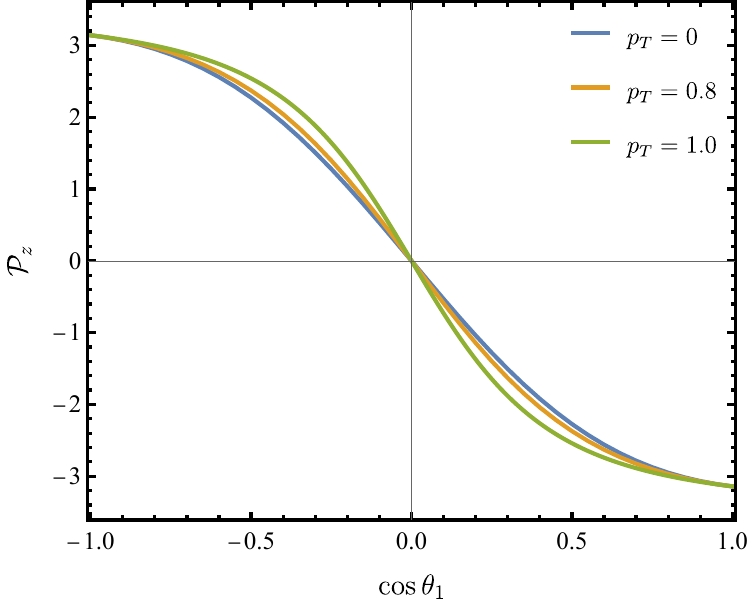}
	\end{center}\vspace{-5ex}
	\caption{The $\cos\theta_1$ distributions of polarization $\cP_z$ for different transverse polarizations of beams. Here $p_L=0.5$ is used. }
	\label{fig:Pz-theta1}
    \vspace{-5mm}
\end{figure}

\begin{figure}[thb]
	\begin{center}
	\includegraphics[scale=.60]{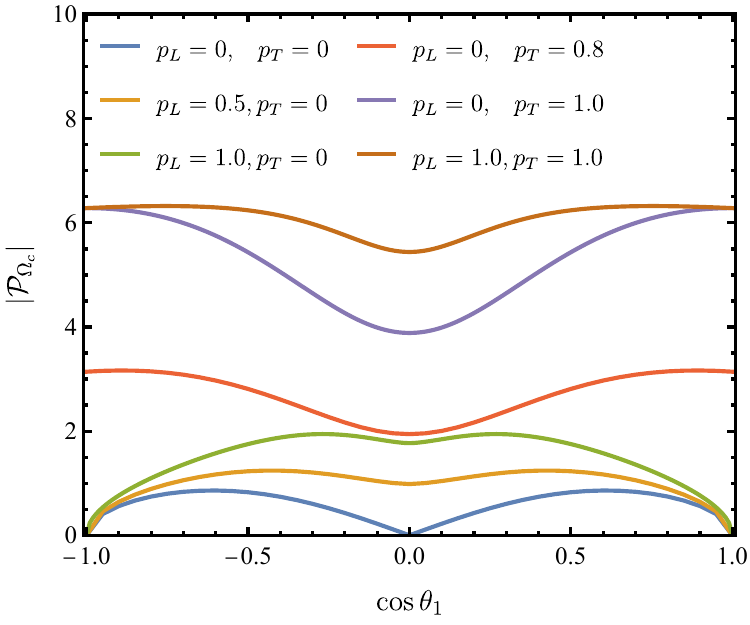}
	\end{center}\vspace{-5ex}
	\caption{Magnitudes of the $\Omega_c$ polarization as a function of $\cos\theta_1$ for different beam polarizations.}
	\label{fig:PAbs-theta1}
\end{figure}

The transverse and longitudinal  polarizations of $\Omega_c$ are 
\begin{align}\label{eq:Px-Py}
\cP_x = &
\frac{\sqrt{1-\alpha_c^2}\sin\theta_1 [(p_L-\bar{p}_L)\cos\Delta_1 -  p_T^2 \sin\Delta_1\sin2\phi_1 ] }{1 + \alpha_c\cos^2\theta_1 + p_T^2 \,\alpha_c  \sin^2\theta_1
\cos2\phi_1},
\nonumber\\
\cP_y = &
\frac{\sqrt{1-\alpha_c^2}\sin\Delta_1\sin\theta_1\cos\theta_1(1- p_L\bar{p}_L -p_T^2\cos2\phi_1) }{1 + \alpha_c\cos^2\theta_1 + p_T^2 \,\alpha_c  \sin^2\theta_1
\cos2\phi_1}
,\nonumber\\
\cP_z = & 
\frac{  (1+\alpha_c) (\bar{p}_L - p_L)\cos\theta_1 }{1 + \alpha_c\cos^2\theta_1 + p_T^2 \,\alpha_c  \sin^2\theta_1
\cos2\phi_1}\,.
\end{align}
These polarizations reduce to the results in Ref.~\cite{Cao:2024tvz} when the incoming beams are polarized transversely, and turn to the expressions in Ref.~\cite{Salone:2022lpt} when electron beam is polarized longitudinally. For continuity and simplicity, we also treat the longitudinal polarization of positron beam to be zero by taking $\bar{p}_L=0$. We have found that the dependences of results below on longitudinal beam polarizations are dominated by the difference $p_L-\bar{p}_L$. The $\cos\theta_1$ distribution of $\cP_y$ can be obtained after integrating out $\phi_1$, and its magnitude is improved by transverse polarization $p_T$ as plotted in Fig.~\ref{fig:Py-theta1}. Since the existence of polarized beams, the polarizations $\cP_x$ and $\cP_z$ are no longer flat, and the $\cos\theta_1$ distribution of $\cP_z$ is shown in Fig.~\ref{fig:Pz-theta1}. The distribution of magnitude $|\cP_{\Omega_c}|=\sqrt{\cP_x^2+\cP_y^2+\cP_z^2}$ is also shown in Fig.~\ref{fig:PAbs-theta1}, which highlights discernible differences associated with different beam polarization configurations. 

\subsection{Two-body decays and joint angular distributions}

The information of $P$ violation of this two-body decay is carried by the SDM of $\Omega^-$, which is given by the relation 
\begin{eqnarray}\label{eq:SDM-spin32}
\rho^{\Omega^-}_{\lambda_{2},\lambda'_{2}}&\propto&\sum_{{\lambda_{1}},\lambda'_{1}}{\rho^{\Omega_c}_{\lambda_{1},\lambda_{1}^{'}}}D^{\frac{1}{2}*}_{\lambda_{1},\lambda_2}(\phi_2,\theta_2)\nonumber\\
	&\times&D^{\frac{1}{2}}_{\lambda_{1}',\lambda_2'}(\phi_2,\theta_2)B_{\lambda_{2}} B_{\lambda_{2}'}\,,
\end{eqnarray}
here the helicity value $\lambda_2$ or $\lambda'_2$ is either $1/2$ or $-1/2$, that means the SDM $\rho^{\Omega^-}$ is also a $2\times 2$ matrix likes \eq{rho-Omec}. Then after summing all combinations and implementing  simplifications, the unpolarized and longitudinal-polarized operators of $\Omega^-$ reduce to 
\begin{align}
\label{eq:P0-Ome}
\cP_0^{\Omega^-}
\! = &
\frac{2\cP_0}{1+\alpha_{\Omega_c}}
\big[
1 +\alpha_{\Omega_c}
\cos\theta_2 \cP_z
\nonumber \\
& \qquad\quad +
\alpha_{\Omega_c}\sin\theta_2 
\left(\cos\phi_2\cP_x 
+ 
\sin\phi_2\cP_y
\right)
\big]\,,
 \\
\cP_0^{\Omega^-}\cP_z^{\Omega^-}
\! = &
\frac{2\cP_0}{1+\alpha_{\Omega_c}}
\big\{
\alpha_{\Omega_c}+
\cos\theta_2 \cP_z
\nonumber \\
& \qquad\qquad +
\sin\theta_2 
\left(\cos\phi_2\cP_x 
+ 
\sin\phi_2\cP_y
\right)
\big\}
\,, 
\end{align}
and the transverse-polarized operators are 
\begin{align}
\cP_0^{\Omega^-}\cP_x^{\Omega^-}
\!= &
\cP_0\sqrt{1-\alpha_{\Omega_c}\over 1+\alpha_{\Omega_c}}
\big\{
\sin\Delta_2
\left( 
\cP_y\cos\phi_2-\cP_x\sin\phi_2 
\right)
\nonumber \\
& - \cos\theta_2 
\left[
\cP_z\sin\theta_2 -
\left( \cP_x\cos\phi_2 +
\cP_y\sin\phi_2 
\right)
\right]
\big\}
, \\
\cP_0^{\Omega^-}\cP_y^{\Omega^-}
\!= &
\cP_0\sqrt{1-\alpha_{\Omega_c}\over 1+\alpha_{\Omega_c}}
\big\{
\cos\Delta_2
\left( 
\cP_y\cos\phi_2-\cP_x\sin\phi_2 
\right)
\nonumber \\
& + \sin\theta_2 
\left[
\cP_z\sin\theta_2 -
\left( \cP_x\cos\phi_2 +
\cP_y\sin\phi_2 
\right)
\right]
\big\}\,.
\label{eq:P0Py-Ome}
\end{align}

\begin{figure}[thb]
	\begin{center}
	\includegraphics[scale=.60]{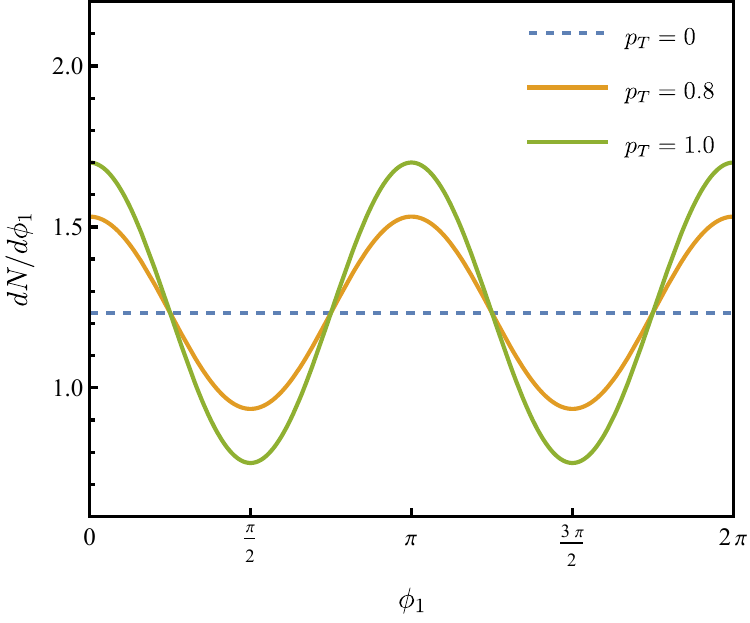}
	\end{center}\vspace{-5ex}
	\caption{The distributions of $\phi_1$ for three different transverse polarizations of beam.  }
	\label{fig:N-theta1}
\end{figure}

\begin{figure}[thb]
	\begin{center}
	\includegraphics[scale=.60]{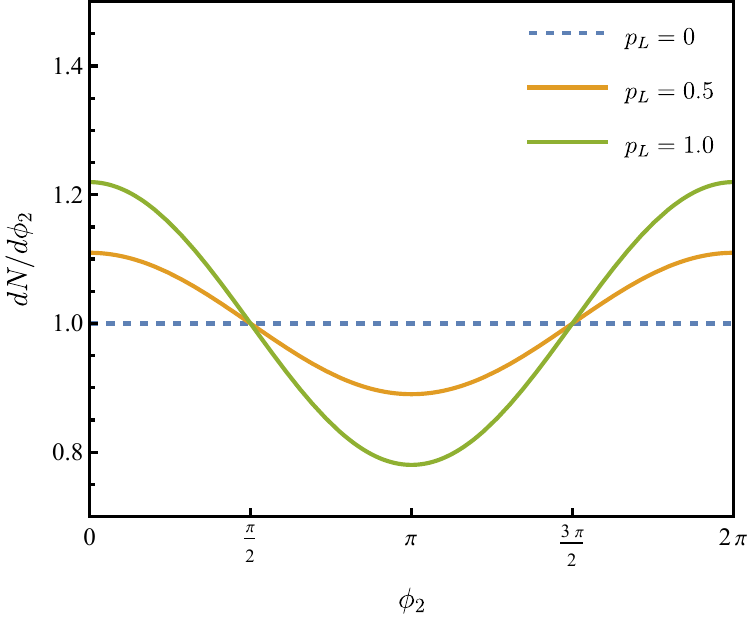}
	\end{center}\vspace{-5ex}
	\caption{The distributions of $\phi_2$ for different longitudinal polarizations of electron beam. The normalization is arbitrary. Here $\alpha_{\Omega_c}=0.7$ is used.}
	\label{fig:N-phi2}
\end{figure}

Finally, all parity information is compiled in the SDM of proton, which is expressed by
\begin{align}
\rho^{p}_{\lambda_{4},\lambda'_{4}}
\propto&
\sum_{{\lambda_{1}},\lambda'_{1}}{\rho^{\Omega_c}_{\lambda_{1},\lambda_{1}^{'}}}
B_{\lambda_{2}} B_{\lambda_{2}'}
F_{\lambda_{3}} F_{\lambda_{3}'}
H_{\lambda_{4}} H_{\lambda_{4}'}
\nonumber
\\\times&
D^{\frac{1}{2}*}_{\lambda_{1},\lambda_2}
(\phi_2,\theta_2)
D^{\frac{1}{2}*}_{\lambda_{2},\lambda_3}(\phi_3,\theta_3)
D^{\frac{1}{2}*}_{\lambda_{3},\lambda_4}(\phi_4,\theta_4)
\nonumber\\\times &
D^{\frac{1}{2}}_{\lambda_{1}',\lambda_2'}(\phi_2,\theta_2)
D^{\frac{1}{2}}_{\lambda_{2}',\lambda_3'}(\phi_3,\theta_3)
D^{\frac{1}{2}}_{\lambda_{3}',\lambda_4'}(\phi_4,\theta_4)
\,.
\end{align}
The polarizations of proton can be obtained using similar expressions of Eqs.~\eqref{eq:P0-Ome}-\eqref{eq:P0Py-Ome} iteratively, and we suppress the explicit expressions here. The joint angular distribution is given by unpolarized section of proton $\cW=\text{Tr}\rho^p = \cP_0^p$. The angular distribution of event number $N$ for each angle is obtained by integrating out other angles, and the nontrivial distributions are
\begin{align}
\label{eq:N-theta1}
\frac{dN}{d\cos\theta_1}
\propto &
1 + \alpha_c\cos^2\theta_1
\,, \\
\label{eq:N-theta3}
\frac{dN}{d\cos\theta_3}
\propto &
1 + \alpha_{\Omega_c}\alpha_{\Omega^-}
\cos\theta_3
\,, \\
\frac{dN}{d\cos\theta_4}
\propto &
1 + \alpha_{\Omega^-}\alpha_{\Lambda}
\cos\theta_4
\,,\\
\label{eq:N-phi1}
\frac{dN}{d\phi_1}
\propto &
1 + \frac{\alpha_c}{3}
\left(
1 + 
2 p_T^2\cos2\phi_1
\right)
\,,\\
\label{eq:N-phi2}
\frac{dN}{d\phi_2}
\propto &
1 + \frac{3\pi^2}{16}\alpha_{\Omega_c}
\frac{\sqrt{1-\alpha_c^2}}{3+\alpha_c} p_L
\cos\Delta_1 \cos\phi_2
\,,\\
\frac{dN}{d\phi_4}
\propto &
1 - \frac{\pi^2}{16}\alpha_{\Omega_c}\alpha_{\Lambda}
\sqrt{1-\alpha^2_{\Omega^-}} \cos(\Delta_3 + \phi_4)\,.
\end{align} 
The distributions of $\theta_2$ and $\phi_3$ are flat. The distributions in \eqs{N-theta1}{N-theta3} are used for fitting the values of parameters $\alpha_c$ and $\alpha_{\Omega_c}$. The distributions of $\phi_1$, $\phi_2$, and $\phi_4$ are no longer flat when the beams are polarized transversely or longitudinally. The plot for $\phi_1$ distribution in \eq{N-phi1} is given in Fig.~\ref{fig:N-theta1}, where the fluctuation is not affected by longitudinal polarization of electron beam but enhanced by the transverse polarization as plotted in Fig.~\ref{fig:N-phi2}. Conversely, the $\phi_2$ distribution is free of the value of $p_T$ but depends on longitudinal polarization $p_L$.

Theoretical predictions for the asymmetry parameter  $\alpha_{\Omega_c}$ are currently lacking. However, for the decays $\Xi_c^+ \to \Xi^0 \pi^+$ and $\Xi_c^0 \to \Xi^- \pi^+$, which involve similar diagrams to those in Fig.~\ref{fig:tree-loop} but with the spectator $s$ quark replaced by a $u$ or $d$ quark, the absolute values of the corresponding asymmetry parameters are predicted to be around $0.95$ and $0.7$ \cite{Zenczykowski:1993jm,Xing:2023dni,Zhong:2022exp,Xing:2024nvg,Zhong:2024qqs,Niu:2025lgt}, respectively. These values are used in the following analysis and should be verified at the STCF via \eq{N-theta3}. On the other hand, no measurement of the relevant parameter for $\Xi_c^+ \to \Xi^0 \pi^+$ is currently available, and it will be accessible at the future STCF.


\begin{figure*}[htb]
	\begin{center}        
\includegraphics[width=.32\textwidth]{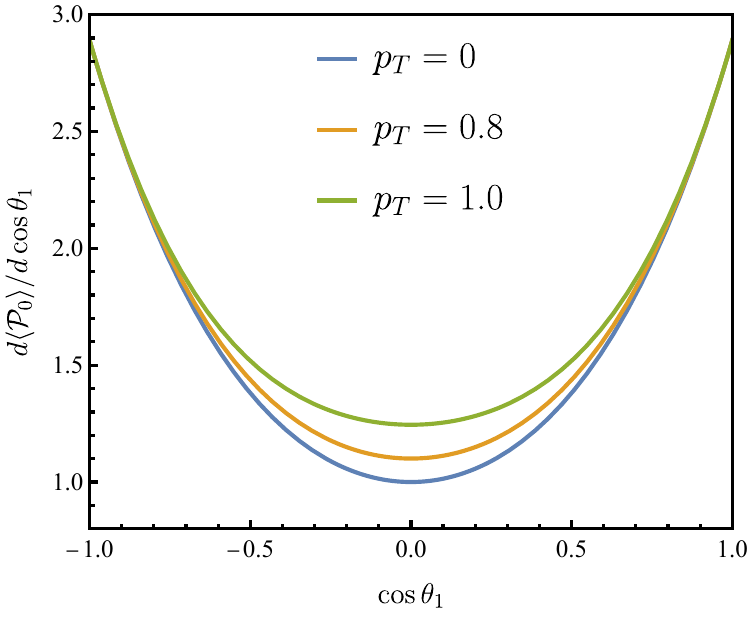}
\includegraphics[width=.32\textwidth]{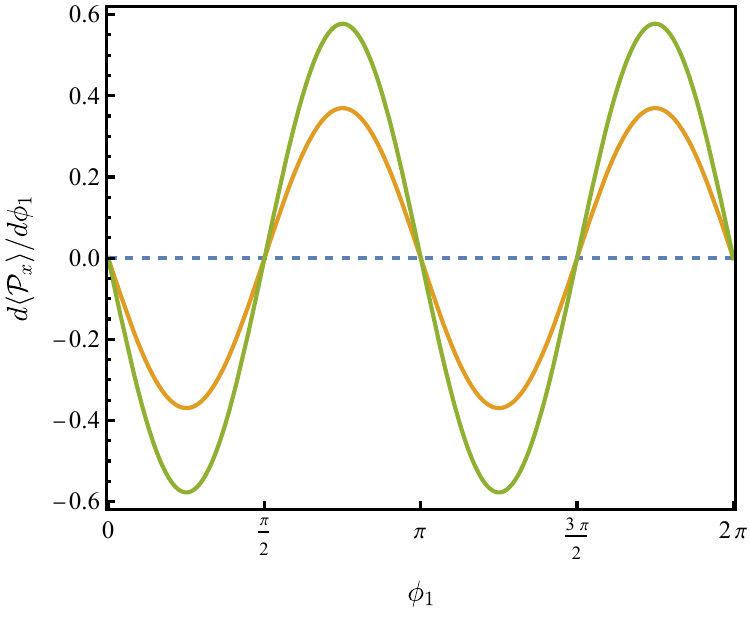}    \includegraphics[width=.32\textwidth]{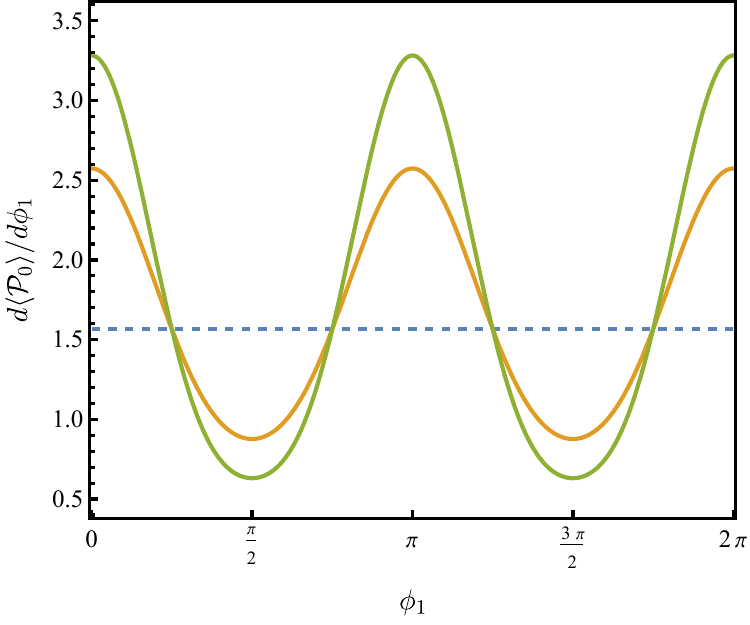}     
	\end{center}\vspace{-4ex}
	\caption{Weighted polarizations of $\Omega_c$ as functions of $\phi_1$ or  $\theta_1$ at different transverse beam polarizations. Here $p_L=0$ is used.}
	\label{fig:P0xy-theta1-phi1}
\end{figure*}

The joint angular distributions of $\theta_1$ and $\phi_1$ are obtained by integrating out all angles of two-body decays, and have the expression:
\begin{align}
\cW(\theta_1,\phi_1)
= &
-\frac{64\pi^3 \cP_0}{(1+\alpha_{\Omega_c})(1+\alpha_{\Omega^-})(1+\alpha_{\Lambda})}
\end{align}
Since the unpolarized operator $\cP_0$ in \eq{P0-photon}
depends on the beam polarization, the statistically measured polarization observables are also affected by the values of $p_L$ and $p_T$. The measurements can be quantified by the weighted polarizations of $\Omega_c$ defined via 
\begin{align}
\langle \cP_i \rangle 
=
\int \cP_{i}\cW(\theta_1,\phi_1)
\,d\cos\theta_1 d\phi_1\,,
\end{align}
where $i\in \{ 0,x,y,z \}$. Explicitly, the expressions of the weighted distributions for $\cos\theta_1$ or $\phi_1$ are 
\begin{align}
\label{eq:ave-P0-theta}
\frac{d\langle \cP_0 \rangle}{d\cos\theta_1}
\propto &\,
1 + 2\alpha_c\cos^2\theta_1 
 + \alpha_c^2
\left(
\cos^4\theta_1 + \frac{p_T^4}{2}\sin^4\theta_1
\right)
\,, \\ 
\frac{d\langle \cP_x \rangle}{d\cos\theta_1}
\propto &\,
p_L\sin\theta_1
\,, \\ 
\frac{d\langle \cP_z \rangle}{d\cos\theta_1}
\propto &\,
p_L \cos\theta_1
\,, \\ 
\frac{d\langle \cP_y \rangle}{d\cos\theta_1}
\propto &\,
\sin\theta_1\cos\theta_1
\,, \\ 
\label{eq:ave-P0-phi}
\frac{d\langle \cP_0 \rangle}{d\phi_1}
\propto & \,
1 + 
\frac{2\alpha_c}{3}
\left( 1 + 2p_T^2\cos2\phi_1  \right)
\nonumber \\ &
\quad + \frac{\alpha_c^2}{5}
(1+2p_T^2\cos2\phi_1)^2
\,, \\ 
\label{eq:ave-Px-phi}
\frac{d\langle \cP_x \rangle}{d\phi_1}
\propto & \,
p_L - p_T^2\tan\Delta_1\sin2\phi_1\,,
\end{align}
where the factor $p_L$ is retained for the distributions that have terms depending on $p_L$. Except $\cos\theta_1$ distribution of $\langle \cP_0 \rangle$ or $\langle \cP_y \rangle$, other distributions are fluctuating only when the beams are polarized. The plots for Eqs.~\eqref{eq:ave-P0-theta}, \eqref{eq:ave-P0-phi} and \eqref{eq:ave-Px-phi} for different beam polarizations are given in Fig.~\ref{fig:P0xy-theta1-phi1}, respectively. The $\phi_1$ distributions for weighted $\cP_y$ and $\cP_z$ are trivially zero. Furthermore, the values of $p_L$ and $\Delta_1$ can be determined by combining \eqs{N-phi2}{ave-Px-phi} under fixed $\alpha_{\Omega_c}$ and $p_T$.

\section{Sensitivity of asymmetric parameters measurements}
Sensitivity estimation is fundamental to experimental design, which links physical measurement precision to data statistics. We refine sensitivity estimation by using the entire decay chain for $\Omega_c$. Our calculations show the expected precision for asymmetric parameters relative to data statistics. This method applies to similar decay processes. As large-scale facilities like STCF~\cite{Achasov:2023gey}, sensitivity estimation is urgently needed to guide the plan for data collection. Currently, the STCF detector and offline software system are in the research and development stage. Therefore, according to the design report~\cite{Achasov:2023gey}, the reconstruction efficiency of this process is estimated to be 30\%. We can discuss whether the facility could confirm or exclude certain theoretical scenarios after considering the impacts of the branching fractions and reconstruction efficiencies.

\begin{figure}[H]
\begin{center}
\begin{overpic}[width=0.35\textwidth, trim=80 60 50 10,angle=0]{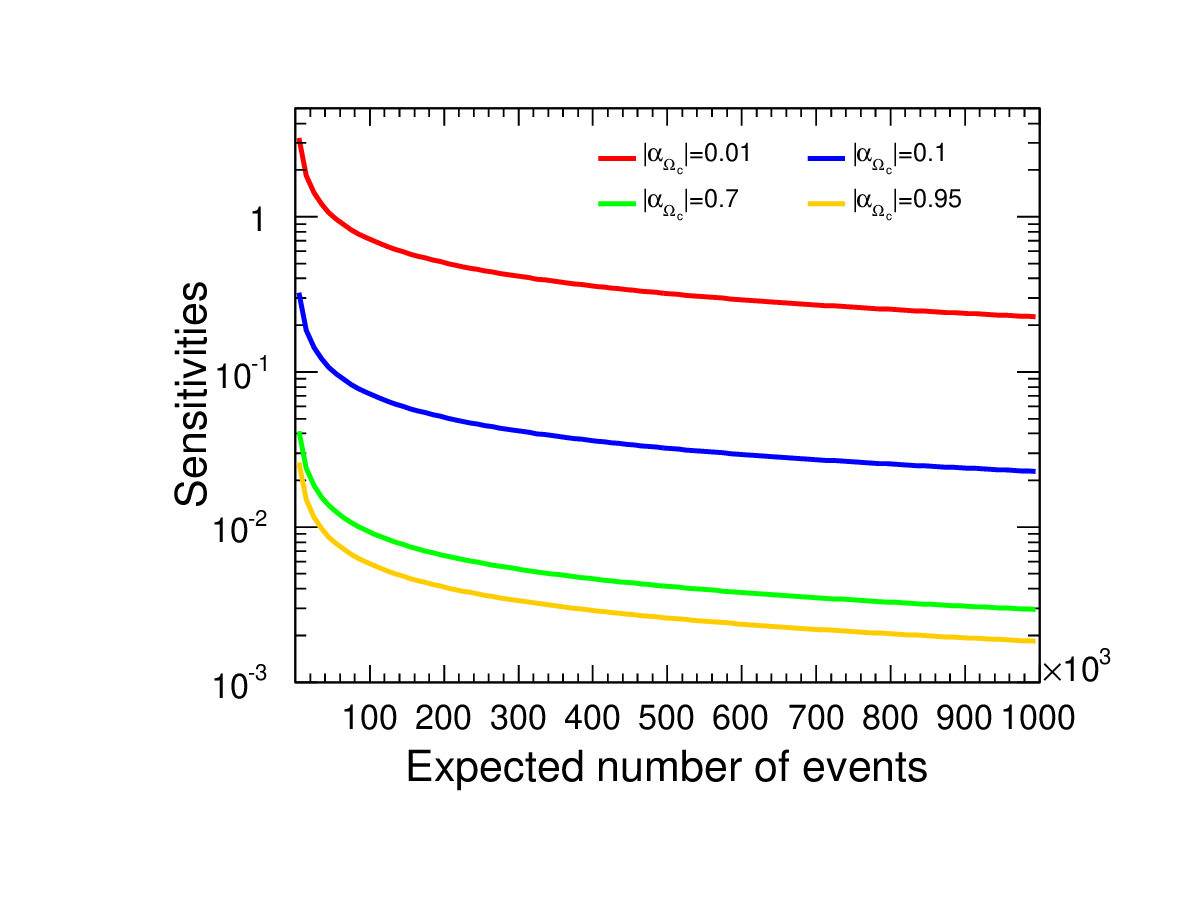}
\end{overpic}
\end{center}
\caption{The $\alpha_{\Omega_c}$ sensitivity distributions relative to signal yields in terms of three different values.
}
\label{fig:delta-alpha}
\end{figure}

Our estimation is based on maximum likelihood method. For the observed data sample of $N$ events, the likelihood function is expressed as~\cite{Han:2019axh}
\begin{align}
L(\theta_i, \phi_i,\alpha_c,\alpha_{\Omega_c},\alpha_{\Omega^-},\alpha_{\Lambda})
=
\prod^N_{i=1} \widetilde{\cW}_j\,,
\end{align}
where $\theta_i$ and $\phi_i$ represent polar angle and azimuth angle. The function depends on relative decay parameters $\alpha_c,\alpha_{\Omega_c},\alpha_{\Omega^-},\alpha_{\Lambda}$ and is computed based on the probability of the $i$-th event $\widetilde{\cW}_j$, whose distribution is normalized. Then the relative uncertainty for estimating statistical sensitivity to parity parameter $\alpha_{\Omega_c}$ is defined as
\begin{align} 
\delta(\alpha_{\Omega_c})
=
\frac{\sqrt{V(\alpha_{\Omega_c})}}{|\alpha_{\alpha_{\Omega_c} }|} \,,
\end{align}
where the inverse of the variance is given by 
\begin{align}
V^{-1}(\alpha_{\Omega_c})
=
N\int \frac{1}{\widetilde{\cW}}
\left[ 
\frac{\partial \widetilde{\cW}}{\partial \alpha_{\Omega_c}}
\right]^2
\prod_{i}^4d\cos\theta_i d\phi_i\,,
\end{align}
where the $N$ denotes the number of observed events~\cite{Han:2019axh}. We consider a set of possible values of $\alpha_{\Omega_c}$ to plot their sensitivities shown in Fig.~\ref{fig:delta-alpha}. At a fixed number of events, the statistical sensitivity is a monotonically decreasing function of the absolute value of the asymmetry parameter.  The phases $\Delta_2=\pi/4$ and $\Delta=\pi/3$ are employed here, and other sets of phase angle values do not significantly affect
the statistical quantities. 

\begin{figure}[H]
\begin{center}
\begin{overpic}[width=0.35\textwidth, trim=80 60 50 10,angle=0]{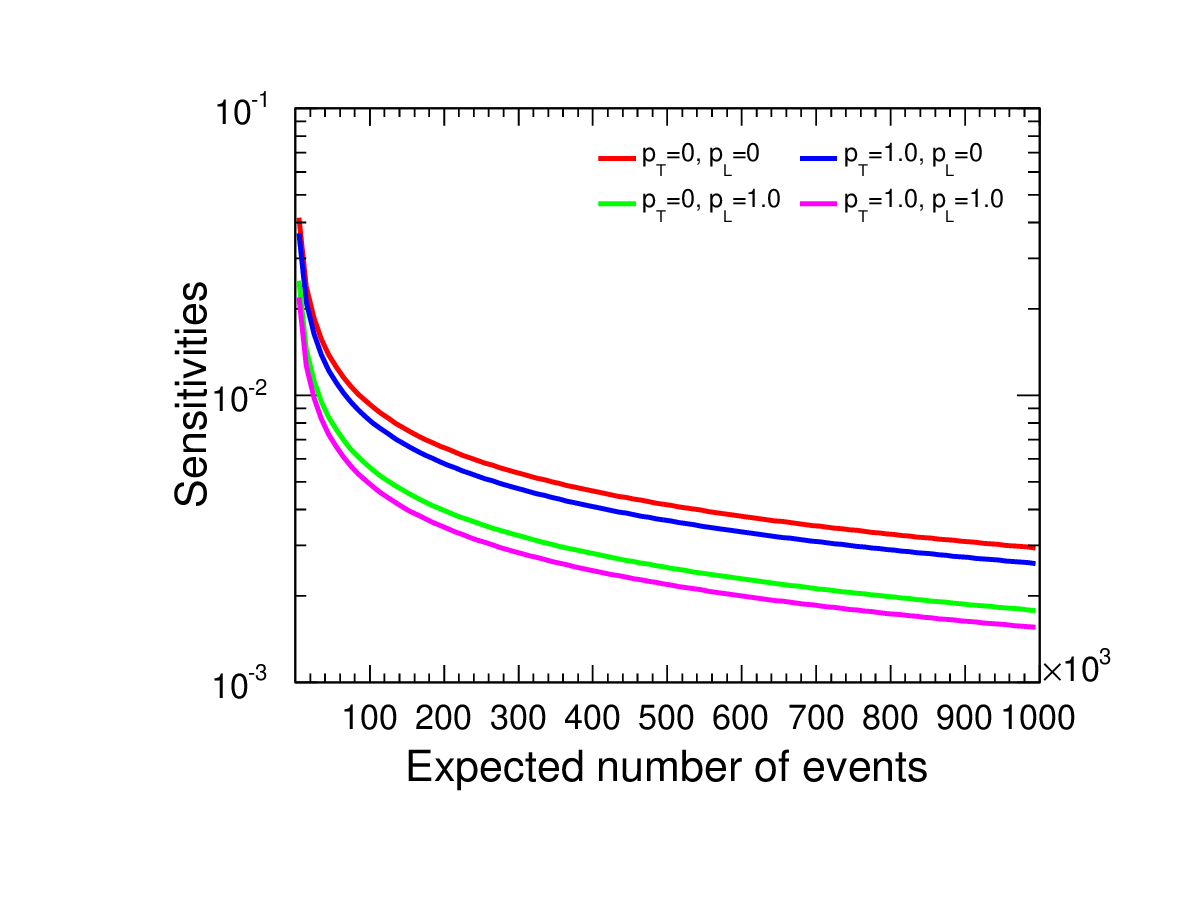}
\end{overpic}
\end{center}
\caption{The $\alpha_{\Omega_c}$ sensitivity distributions relative to signal yields in terms of different beam polarizations.
}
\label{fig:delta-alpha-pol}
\end{figure}

We are particularly interested in the dependence of statistical sensitivities on asymmetry parameters and beam
polarizations. The distributions of $\delta(\alpha_{\Omega_c})$ for four choices of the beam polarizations are plotted in Fig.~\ref{fig:delta-alpha-pol} with $\alpha_{\Omega_c}$ fixed at 0.7. The sensitivity improves as its value decreases by approximately $10\%$ when the beam is transversely polarized at $p_T=1$ compared to an unpolarized scenario, and the value decreases by approximately $34\%$ when the beam is longitudinally polarized at $p_L=1$. 
\vspace{-1mm}
\vspace{-8mm}

\begin{figure}[H]
\begin{center}
\begin{overpic}[width=0.35\textwidth, trim=80 60 50 10,angle=0]{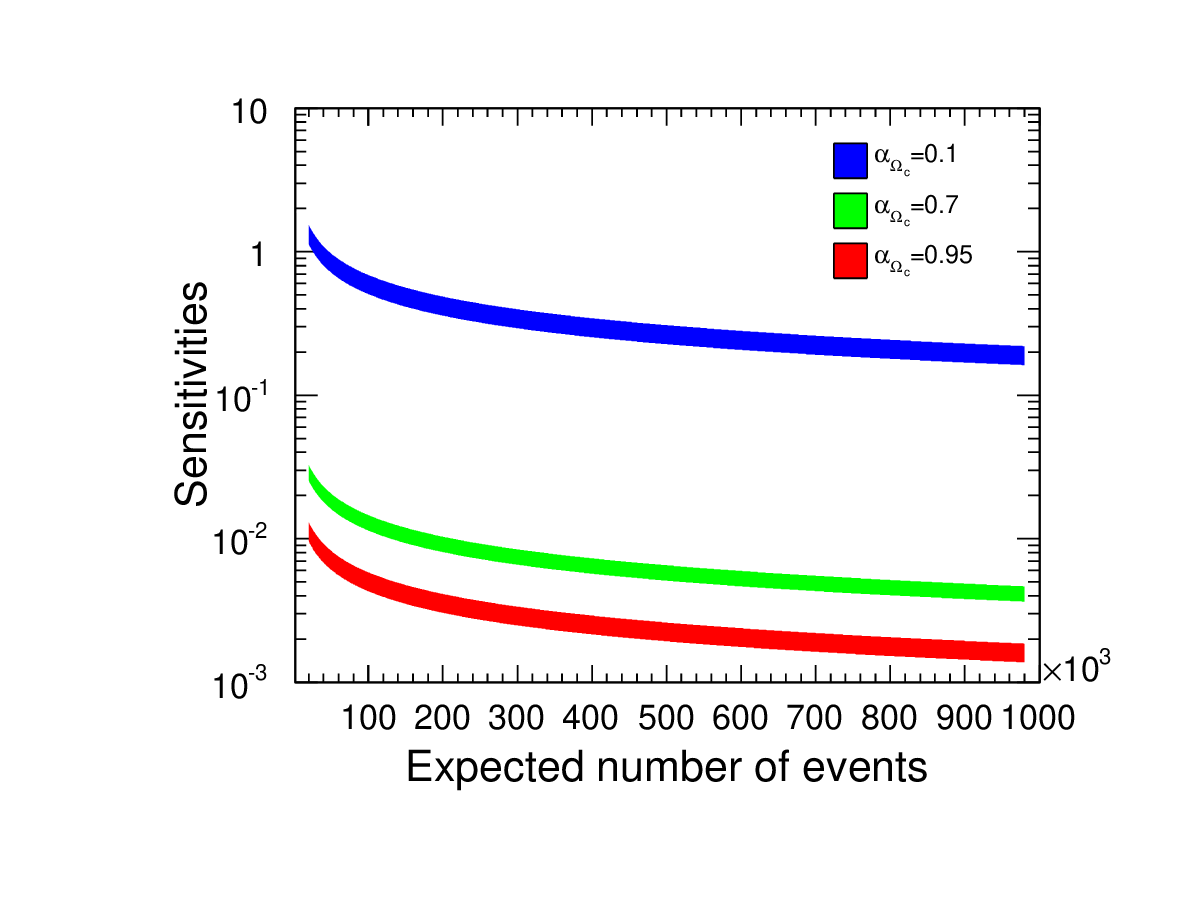}
\end{overpic}
\end{center}
\caption{The $\mathcal{A}_{CP}$ sensitivity distributions relative to signal yields in terms of two different values of $\alpha_{\Omega_c}$. The uncertainty from $CP$ parameter is $|\mathcal{A}_{CP}|<0.076$.
}
\label{fig:delta-ACP}
\end{figure}

\begin{figure}[H]
\begin{center}
\begin{overpic}[width=0.35\textwidth, trim=80 60 50 10,angle=0]{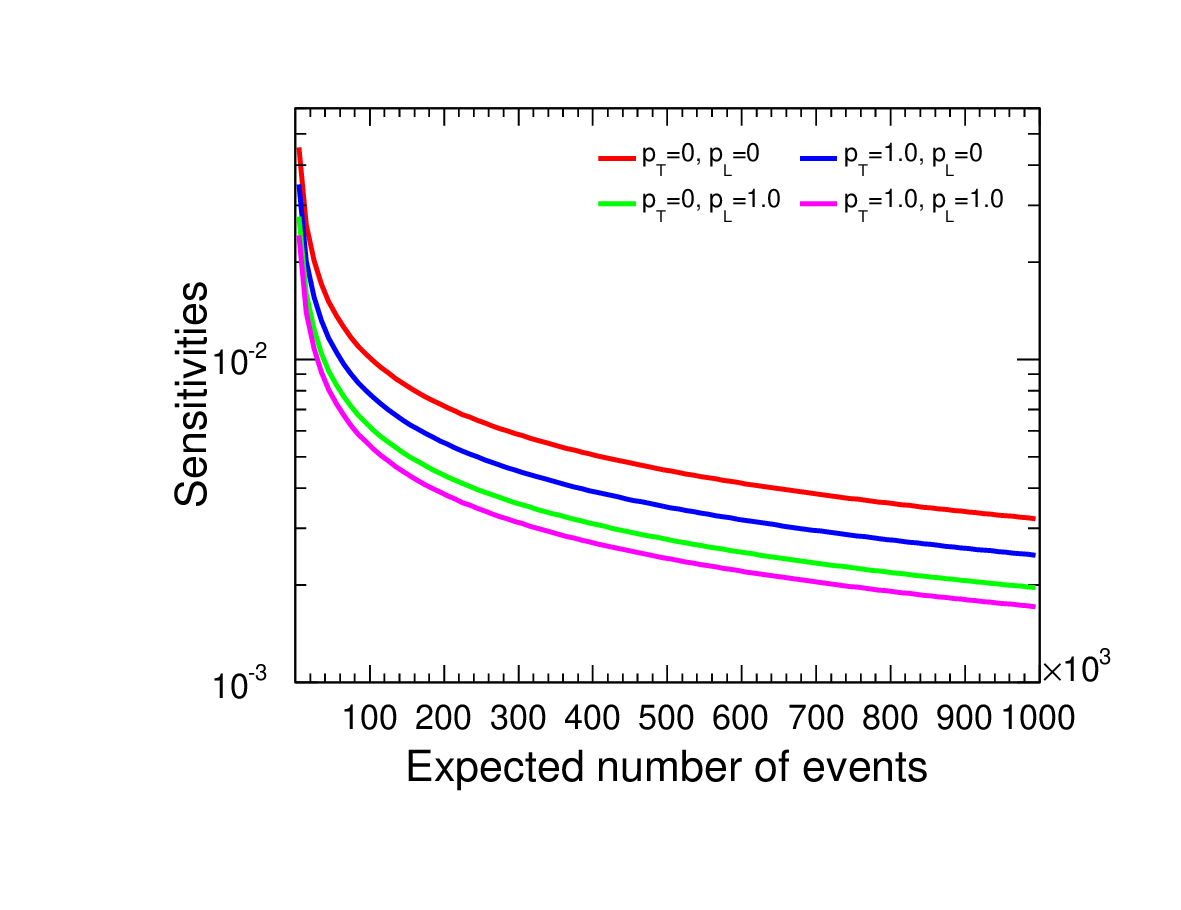}
\end{overpic}
\end{center}
\caption{The $\mathcal{A}_{CP}$ sensitivity distributions relative to signal yields in terms of different beam polarizations.
}
\label{fig:delta-ACP-pol}
\end{figure}

Indeed for identifying significance of $CP$ violations, the statistical sensitivity of $\mathcal{A}_{CP}$ in \eq{Acp} can be estimated if $\alpha_{\Omega_c}$ and $\bar\alpha_{\bar{\Omega}_c}$ are considered as non-correlation, via error propagation formula:
\begin{align}
\label{eq:delta-Acp}
\delta(\mathcal{A}_{CP})
= 
\frac{2\sqrt{\alpha_{\Omega_c}^2\delta(\bar{\alpha}_{\bar{\Omega}_c})^2 + \bar{\alpha}_{\bar{\Omega}_c}^2 \delta(\alpha_{\Omega_c})^2 } }{(\alpha_{\Omega_c} - \bar{\alpha}_{\bar{\Omega}_c})^2}\,,
\end{align}
where $\delta(\bar{\alpha}_{\bar{\Omega}_c})$ is the sensitivity for $\bar{\Omega}_c$ system. The plots for sensitivity $\delta(\mathcal{A}_{CP})$ is shown in Fig.~\ref{fig:delta-ACP} with choosing three different values of $\alpha_{\Omega_c}$. The sensitivity and the $CP$ parameter are positively correlated when other variables are fixed, and the bands are used to display uncertainty from variation $|\mathcal{A}_{CP}|<0.076$. Though the direct $CP$ violation from weak interactions for decays of charm baryons is of order $10^{-4}$ or less, we adopt the upper limit $0.076$ from experiment~\cite{Belle:2021crz} to involve possible enhancements from other mechanisms. The value of $\mathcal{A}_{CP}$ can be obtained by combining \eqs{Acp}{N-theta3}, associated with determined $\alpha_{\bar{\Omega}_c}$ from the conjugation decay. The $\delta(\mathcal{A}_{CP})$ for four choices of the beam polarizations are plotted in Fig.~\ref{fig:delta-ACP-pol} with $\alpha_{\Omega_c}$ fixed at 0.7. They exhibit similar line shapes and evolutionary trends in Fig.~\ref{fig:delta-alpha-pol}.

The expected number of observed signal events each year is estimated, assuming that the cross section of $e^+e^-\rightarrow \Omega_c\bar{\Omega}_c$ is approximately equal to that of $e^+e^-\rightarrow \Lambda_c^+\bar{\Lambda}_c^-$~\cite{BESIII:2023rwv,BESIII:2024xgl}.
The branching fractions of intermediate processes and detection efficiency have been considered in detailed calculation.
The branching fraction of $\Omega_c\rightarrow\Omega^-\pi^+$ decay has been computed in Ref.~\cite{Cheng:1996cs,Hsiao:2020gtc,Wang:2022zja,Zeng:2024yiv} and the value $4.2\%$~\cite{Cheng:1996cs} is chosen in our analysis. The branching fractions of $\Omega^-\rightarrow\Lambda K^+$ and $\Lambda\rightarrow p \pi^-$ are taken from Particle Data Group~\cite{ParticleDataGroup:2024cfk}. The detection efficiency of the signal channel is roughly estimated to $30\%$. Therefore, the observed $\Omega_c\rightarrow\Omega^-\pi^+\rightarrow\Lambda K^+\pi^+\rightarrow p\pi^-K^+\pi^+$ yields are expected to $\sim 0.37$ million at STCF. Approximate values for $\alpha_{\Omega_c}$, inferred from the predicted parameters of $\Xi_c^+$ and $\Xi_c^0$~\cite{Niu:2025lgt}, can be tested at the STCF with a precision of $0.48\%$ and $0.34\%$ for $|\alpha_{\Omega_c}| = 0.7$ and $0.95$, respectively, as shown in Figure~\ref{fig:delta-alpha}.
The expected $CP$ violation in $\Omega_c\rightarrow\Omega^-\pi^+$ decay from weak interactions is on the order of $10^{-9}$ to $10^{-10}$. Therefore, it is estimated that at least $2.0\times10^{18}$ $\Omega_c\bar{\Omega}_c$ events are required to observe $CP$ violation, under the assumption of $\alpha_{\Omega_c} =0.95$. We can achieve precise measurement of decay parameter and $CP$ parameter at STCF, but the data sample is not sufficient to observe $CP$ violation unless some new physical contributions are made.
The detailed statistical sensitivities $\delta(\alpha_{\Omega_c})$ and $\delta(\mathcal{A}_{CP})$ for different parameter assumptions are listed in Tab.~\ref{tab:sensitivity}.

\begin{table}[H]
\footnotesize
\caption{Statistical sensitivities $\delta_{\alpha_{\Omega_c}}$ and $\delta_{\mathcal{A}_{CP}}$ for different parameter assumptions based on expected 0.37 million events. Here, $\mathcal{A}_{CP}$ is set to 0.001.} 
\label{tab:sensitivity}
\begin{center}
\begin{tabular}{ccc}
\hline \hline 
 Assumed parameters & \quad $\delta_{\alpha_{\Omega_c}}(\%)$
& \quad $\delta_{\mathcal{A}_{CP}}(\%)$ \\
\hline
$\alpha_{\Omega_c}= 0.7$, $p_T= 0$,~~ $p_L=0$~~ & 0.48 & 0.49
\\ 
$\alpha_{\Omega_c}= 0.95$, $p_T= 0$,~~ $p_L=0~~$ & 0.34 & 0.27
\\ 
$\alpha_{\Omega_c}= 0.95$, $p_T= 1.0$, $p_L=0$~~ & 0.28 & 0.22
\\ 
$\alpha_{\Omega_c}= 0.95$, $p_T= 0$,~~ $p_L=1.0$ & 0.16 & 0.13
\\
$\alpha_{\Omega_c}= 0.95$, $p_T= 1.0$, $p_L=1.0$ & 0.14 & 0.10
\\ \hline \hline 
\end{tabular}
\end{center}
\end{table}

\section{summary}
The beam polarization can affect the transverse and longitudinal polarization distributions of $\Omega_c$, as well as the angular distributions of $\theta_1$ and $\phi_1$ in the center-of-mass frame of the electron-positron system. By measuring the angular distribution of the $\Omega_c$ and particles from $\Omega_c$ decay, we can infer the contributions of the transverse and longitudinal polarization of the electron-positron beam. 
Although both longitudinal and transverse beam polarizations can improve the precision of decay parameters and $CP$ parameters, the longitudinal polarization exhibits greater sensitivity than the transverse polarization. 

Currently, experiments have achieved preliminary control of beam polarization. For future studies, particularly in precision $CP$ measurements, achieving controllable transverse and longitudinal polarization with high polarization ratio will be crucial. 
In this study, we evaluate the sensitivities of the asymmetry parameters and $CP$ violation in the decay $\Omega_c \to \Omega^- \pi^+$ under various data sample sizes and beam polarization conditions. If the decay parameter itself is larger, the sensitivity will be better, which means that an accurate result can be measured experimentally. The polarization-dependent angular distributions and sensitivity estimates developed in this work can be used to enhance the verification of various theoretical predictions for related decays such as $\Xi_c^+ \to \Xi^0 \pi^+$ and $\Xi_c^0 \to \Xi^- \pi^+$ at the STCF. This work represents the first investigation of $P$ and $CP$ violation in the two-body decay of $\Omega_c$, providing essential theoretical support for future experiments, such as STCF.

\end{document}